\documentclass[a4paper,11pt]{article}
\usepackage{pos}
\usepackage[shortlabels]{enumitem}
\usepackage{siunitx}
\usepackage{tikz}
\usepackage{xspace}

\newcommand{\peanuts}{\texttt{PeANuTS}\xspace}

\title{Identification of time-correlated neutrino clusters in populations of astrophysical transient sources}
\ShortTitle{Time-correlated neutrino clusters in populations of transient sources}

\author*[a]{Mathieu Lamoureux}
\author[a]{Gwenhaël de Wasseige}
\affiliation[a]{Centre for Cosmology, Particle Physics and Phenomenology - CP3, \\ Université Catholique de Louvain, B-1348 Louvain-la-Neuve, Belgium}

\emailAdd{mathieu.lamoureux@uclouvain.be}

\abstract{The detection of astrophysical neutrinos from transient sources can help to understand the origin of the neutrino diffuse flux and to constrain the underlying production mechanisms. In particular, proton-neutron collisions may produce GeV neutrinos. However, at these energies, neutrino data from large water Cherenkov telescopes, like KM3NeT and IceCube, are dominated by the well-known atmospheric neutrino flux. It is then necessary to identify a sub-dominant component due to an astrophysical emission based on time correlation across messengers. The contribution covers several methods to search for such a signal in short time windows centered on observed transient sources, including a novel approach based on the distribution of time differences. Their performance is compared in the context of subpopulations of astrophysical sources that may show prompt or delayed neutrino emissions. The outlook for the usage of such techniques in actual analyses is also presented.}

\ConferenceLogo{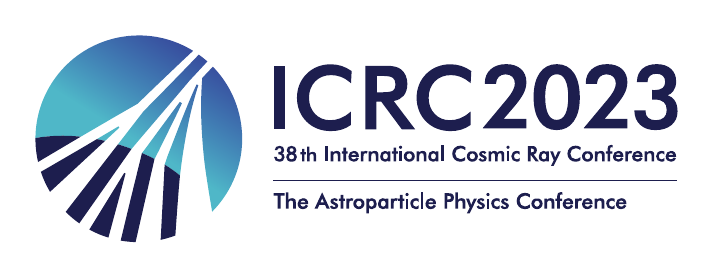}

\FullConference{%
The 38th International Cosmic Ray Conference (ICRC2023)\\
  26 July -- 3 August, 2023\\
  Nagoya, Japan}

\begin{document}

\maketitle

\section{Introduction}

Since the discovery of high-energy astrophysical neutrinos from the IceCube Collaboration in 2013~\cite{IceCube:2013low}, neutrino telescopes have continuously been monitoring the sky in search of the corresponding sources. Multi-messenger approaches have allowed identifying some candidates. In 2017, a high-energy neutrino from IceCube has been detected from the direction of the blazar TXS 0506+056, in time coincidence with a flare in gamma rays~\cite{IceCube:2018cha}. More recently, several tidal disruption events have also been associated with neutrinos in IceCube~\cite{Stein:2020xhk,Reusch:2021ztx}. 

One key feature to clearly associate neutrinos with an astrophysical source is to rely on the spatial and temporal coincidences of the signal events. For transient sources, emitting on a short time scale, such a method is very efficient at reducing the contamination from background neutrino events, such as the ones originating from atmospheric neutrinos. 

In the case of the IceCube GeV sample described in~\cite{IceCube:2021jwt,IceCube:2021ddq}, no direction reconstruction is available at the time of writing, so it is only possible to apply time cuts. The expected background rate is at the level of \qty{20}{\milli\hertz}. This sample is sensitive to all-flavor neutrinos at energies ranging from $0.5$ to \qty{100}{\giga\electronvolt}, making it a nice probe for potential GeV neutrino emission.

This work focuses on the search for a short neutrino signal from sub-populations of astrophysical transient sources. The exact delay $\Delta t$ of the neutrino emission with respect to the time $t_0$ of the detection with another messenger is not known. Other studies often employ a conservative \qty{1000}{\second} time window centered on $t_0$~\cite{Baret:2011tk}. However, in the case of the IceCube GeV sample, this corresponds to $\sim 20$ expected background events, limiting the discovery potential for sources that may emit a flux corresponding to only a few detectable neutrinos.

As long as no clear detection is reported, the value of $\Delta t$ is not known a priori. One may expect it to be of the same order within a given population and for a given emission mechanism. It is then possible to reduce the impact of background events by looking for a coherent excess of neutrinos with the same delay with respect to $t_0$ over a set of transients. Additionally, several timings may be relevant for a given source type, such as precursor, prompt, and delayed emissions.

\begin{figure}[hbtp]
    \centering
    \includegraphics[width=\linewidth]{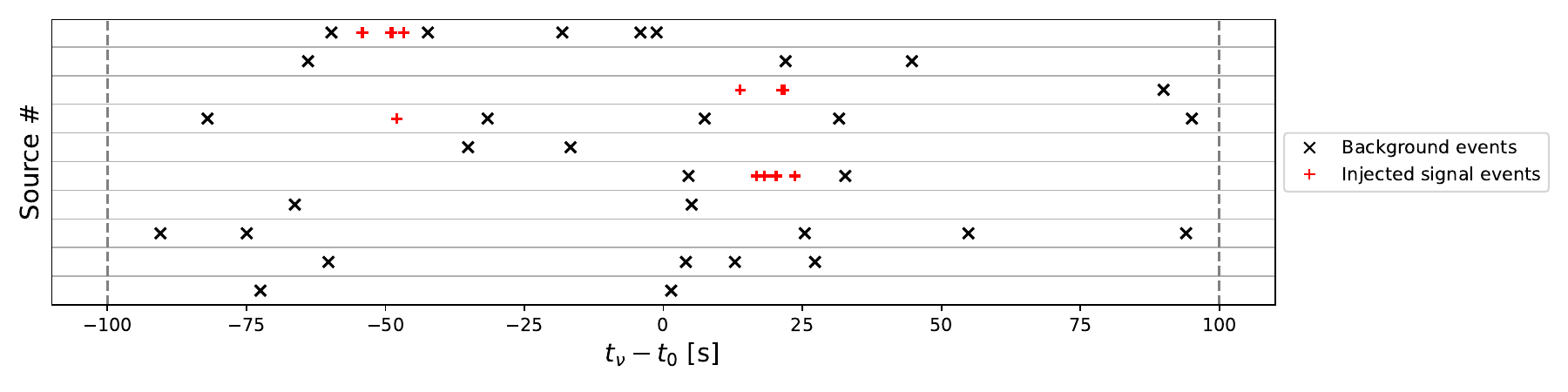}
    \caption{Illustration of the time distribution of neutrino candidate events in toy experiments within $t_0 \pm \qty{100}{\second}$. Each row corresponds to one realization. The black crosses indicate background events with a Poisson rate $r_{\rm bkg} = \qty{20}{\milli\hertz}$ (illustrative of the IceCube GeV sample). The red pluses are injected signal events at $\Delta_t = \qty{-50}{\second}$ and $\Delta_t = \qty{20}{\second}$, with a width of \qty{2}{\second}.}
    \label{fig:ex_timing}
\end{figure}

The goal of this work is to identify subpopulations in a set that may contain sources with no detected signal neutrinos, sources with a signal at $\Delta t$, or sources with a signal at a different timing. The \autoref{fig:ex_timing} shows an illustration of the time distribution of detected events in toy experiments.

\section{Method}
\label{sec:method}

The main ingredient of the search described here is the relative time $\{t_i\}$ of all detected neutrino candidates with respect to the considered transient time $t_0$. For convenience, the analysis is restricted to a time window $[T_{\min}, T_{\max}]$ large enough to contain any kind of signal. To be consistent with other searches, $T_{\min}, T_{\max} = \qty{-500}{\second},\, \qty{500}{\second}$ will be employed.

\subsection{MLE approach}
\label{sec:method:mle}

A usual technique to identify a flaring excess is to define the following likelihood:
\begin{equation}
    \mathcal{L}(N, \{t_i\}; n_s, \Delta t, \sigma_t) = \prod_{i=1}^{N} \left[ \dfrac{n_s}{N} \mathcal{S}(t_i, \Delta t, \sigma_t) + \left(1-\dfrac{n_s}{N}\right) \mathcal{B}(t_i) \right],
\end{equation}
where $N$ is the total number of events, $\{t_i\}$ are their detection times, $n_s$ is the expected number of signal events, $\Delta t, \sigma_t$ is the signal central time and its width, and $\mathcal{S} (\mathcal{B}$) is the signal (background) pdf. In the following, $\mathcal{S}$ is a Gaussian distribution with $\mu = \Delta t$, $\sigma = \sigma_t$, and $\mathcal{B}(t) = 1/(T_{\max} - T_{\min})$. It is then possible to write the Maximum Likelihood Estimator
\begin{equation}
    \textrm{MLE} = 2\log\left( \dfrac{\mathcal{L}(N, \{t_i\}; \hat{n_s}, \hat{\Delta t}, \hat{\sigma_t})}{\mathcal{L}(N, \{t_i\}; n_s=0)} \right),
\end{equation}
where the hatted terms are the values maximizing the likelihood, while the denominator is the background-only likelihood. The maximization is done with the \texttt{iminuit} package~\cite{iminuit}.

\begin{figure}[hbtp]
    \centering
    \includegraphics[width=\linewidth]{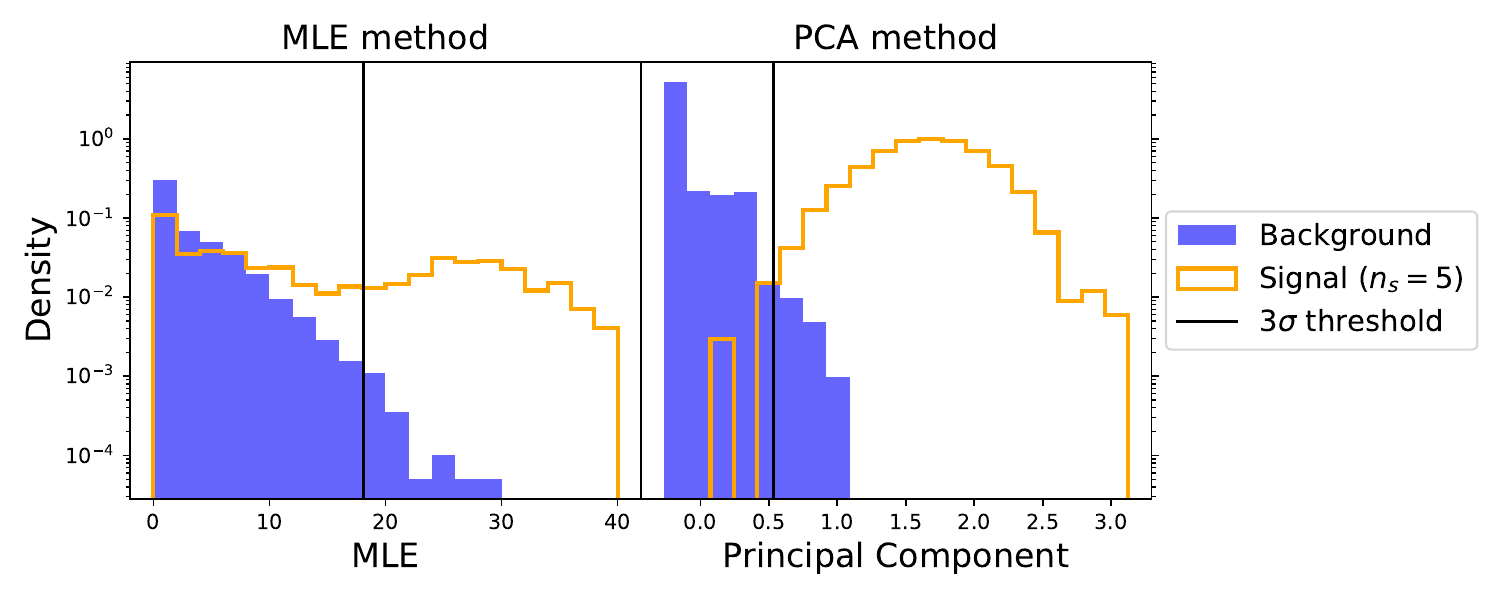}
    \caption{Illustration of the MLE (left) and PCA (right) methods. The blue histogram corresponds to the expected distribution for background events with constant rate $r_{\rm bkg} = \qty{20}{\milli\hertz}$. The orange histogram is the distribution for an injected signal in the search window, with the number of signal events following a Poisson distribution with $\lambda=5$ and a time distribution following a Gaussian distribution with $\sigma=\qty{2}{\second}$. The black line indicates the $3\sigma$ threshold derived from the background distribution. In this example, there are only two sub-populations, those with only background and those with an injected signal, so the PCA threshold is defined by looking at the first component and using the same strategy as for MLE.}
    \label{fig:stdmethod}
\end{figure}

The background distribution of MLE $P_B(\textrm{MLE})$ is obtained using background-only experiments, as illustrated in \autoref{fig:stdmethod}. The significance of a measurement with $\textrm{MLE} = \textrm{MLE}_{\rm observed}$ is then characterized by the p-value $p = \int_{\textrm{MLE}_{\rm observed}}^{\infty} P_B(\textrm{MLE}) {\rm d}\textrm{MLE}$.

For each transient, one may then compute $p$ and consider an observation to be significant if $p<\num{3e-3}$, this would correspond to a $\sim 3\sigma$ deviation from the background expectation. The values of $\hat{n_s}$, $\hat{\Delta t}$, $\hat{\sigma_t}$ may also be recovered to characterize the potential signal. The selection efficiency of the approach can be evaluated using simulations as shown in \autoref{fig:stdmethod}.

Although such a method has proven its liability for such flare searches, it does not take into account the correlations of the signal times across several time windows but only performs independent probes for individual transient.

\subsection{PCA approach}
\label{sec:method:pca}

As the objective is to separate the full dataset into different subpopulations based on the presence or the absence of a neutrino signal, clustering techniques are appropriate for the task. Principal Components Analysis (PCA) is well suited to reduce the dimensionality of the problem before performing clustering, as initially presented in~\cite{deWasseige:2021euk}.

The data is summarised by the set $\{t_i\}$ of observed relative times. To make it suitable for PCA, they are binned in $K$ slices such that $n_j = \sum_{i=1}^{N} H(t_i - \tau_j) H(\tau_{j+1} - t_i)$ for $j=\{0, \ldots, K-1\}$, where $H$ is the Heavyside step function, and $\tau_j = T_{\min} + j \frac{T_{\max} - T_{\min}}{K}$. The overall dataset has then a dimension $M \times K$, where $M$ is the number of analyzed time windows. 

PCA is applied to reduce the number of features to one or two using \texttt{scikit-learn} existing methods~\cite{Pedregosa:2011ork}. This (these) feature(s) may then be used to easily cut between background events and potentially interesting time windows. To evaluate the performance of the search, it is necessary to make use of a partially-labeled dataset. The following procedure has been applied:
\begin{enumerate}
    \item A dataset of pure background events $\mathcal{D}_{\rm bkg}$ is formed. In the following example, this will consist in pseudo-experiments where no signal is injected but only the expected background. For actual measurements, one may consider time windows in real data where no transients were detected.
    \item $\mathcal{D}_{\rm bkg}$ is appended to the dataset of time windows of interest $\mathcal{D}_{\rm search}$ (corresponding to actual data or to simulations with some injected signal): $\mathcal{D} = \mathcal{D}_{\rm bkg} \cup \mathcal{D}_{\rm search}$.
    \item PCA reduction is performed on the full dataset: $\mathcal{D} \to \mathcal{D}^\star = f(\mathcal{D})$.
    \item The distribution of the remaining features is plotted only for the samples originating from $\mathcal{D}_{\rm bkg}$. It is then possible to define a threshold such that any observation exceeding this threshold would correspond to a $3\sigma$ excess.
    \item The observed $f(\mathcal{D}_{\rm search})$ are compared with this threshold to search for any significant event.
\end{enumerate}

A simple example is shown in \autoref{fig:stdmethod}. The performance will be compared with other methods in \autoref{sec:results}. As compared to the likelihood approach, the PCA approach properly covers the needs for the search of subpopulations with similar timings with respect to the detected transient.

\subsection{\peanuts method}
\label{sec:method:peanuts}

Going back to the basics, the current search consists in searching for consistent deviations from the Poisson expectation (that correctly described the background). The available database consists of a succession of neutrino candidate events characterized by their relative detection times with respect to the considered transient source detection time, across time windows covering a population of such sources.

When considering the event times as our only input, a divergence from the Poisson expectation may be identified by plotting the inter-arrival time $\delta t_i = t_{i}-t_{i-1}$ distribution and looking for deviations from an exponential distribution $p(x) = 1/r \exp(-x/r)$, where $r$ is the expected rate. This also applies if one computes the delay between a random time and the next (or previous) event.

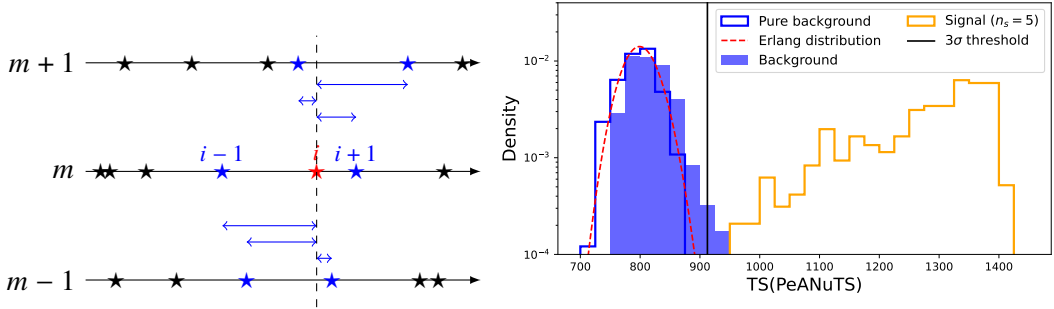
\begin{figure}[hbtp]
    \centering
    \begin{tikzpicture}[xscale=0.4, yscale=1.8]
        \draw[-latex] (0,0) node[left] {$m-1$} -- (13,0);
        \draw[-latex] (0,0.8) node[left] {$m$} -- (13,0.8);
        \draw[-latex] (0,1.6) node[left] {$m+1$} -- (13,1.6);
        
        \node at (1, 0) {$\star$};
        \node at (3, 0) {$\star$};
        \node[blue] at (5.3, 0) {$\star$};
        \node[blue] at (8.1, 0) {$\star$};
        \node at (11, 0) {$\star$};
        \node at (11.6, 0) {$\star$};
        
        \node at (0.5, 0.8) {$\star$};
        \node at (0.8, 0.8) {$\star$};
        \node at (2, 0.8) {$\star$};
        \node[blue] (mim) at (4.5, 0.8) {$\star$};
        \node[red] (mi) at (7.6, 0.8) {$\star$};
        \node[blue] (mip) at (8.9, 0.8) {$\star$};
        \node at (11.8, 0.8) {$\star$};

        \node[scale=.8, red] at (mi.north) {$i$};
        \node[scale=.8, blue] at (mim.north) {$i-1$};
        \node[scale=.8, blue] at (mip.north) {$i+1$};
        
        \node at (1.3, 1.6) {$\star$};
        \node at (3.5, 1.6) {$\star$};
        \node at (6, 1.6) {$\star$};
        \node[blue] at (7.0, 1.6) {$\star$};
        \node[blue] at (10.6, 1.6) {$\star$};
        \node at (12.4, 1.6) {$\star$};
        
        \draw[blue,<->] (7.6, 1.20)--(8.9, 1.20);
        \draw[blue,<->] (7.6, 1.32)--(7.0, 1.32);
        \draw[blue,<->] (7.6, 1.44)--(10.6, 1.44);
        \draw[blue,<->] (7.6, 0.40)--(4.5, 0.40);
        \draw[blue,<->] (7.6, 0.28)--(5.3, 0.28);
        \draw[blue,<->] (7.6, 0.16)--(8.1, 0.16);

        \draw[dashed] (7.6, -0.2)--(7.6, 1.8);
    \end{tikzpicture}
    \includegraphics[width=0.5\linewidth]{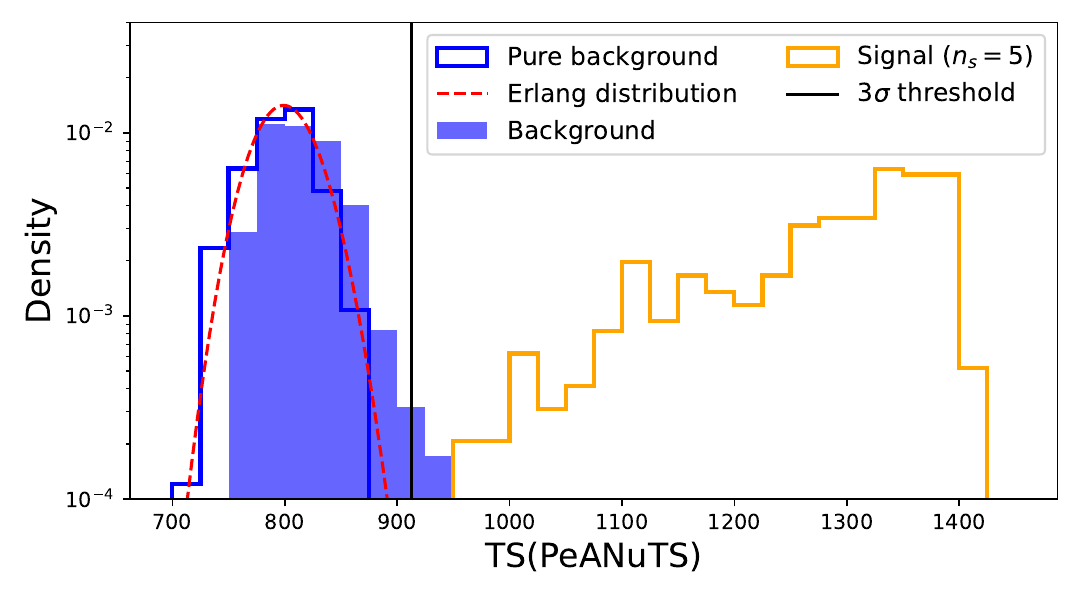}
    \caption{Left: Illustration of the definition of the $\Delta t_{m,i}^{n,-}$ and $\Delta t_{m,i}^{n,+}$ values. Right: The blue step histogram corresponds to pure background, well fitted by an Erlang distribution shown in dashed red. The blue-filled and orange step histograms are the background and signal distributions with the same inputs as in \autoref{fig:stdmethod}.}
    \label{fig:newmethod}
\end{figure}

In the following, $M$ time windows of interest (for instance $\pm \qty{500}{\second}$ around each selected transient source) are considered and the time of each event is noted $t_{m, i}$, where $m$ is the label of the time window and $i$ is indexing the event in the time window. For each event $(m, i)$, one may compute
\begin{align*}
    \Delta t_{m,i}^{n,-} &= \left\{ 
        \begin{array}{l}
        t_{m,i} - t_{m,i-1}\text{, for }n = m, \\
        \min\left[ t_{m,i} - t_{n,j} \right]_{t_{n,j} < t_{m,i}}\text{, for }n \neq m,
        \end{array}
    \right. \\
    \text{and } \Delta t_{m,i}^{n,+} &= \left\{ 
        \begin{array}{l}
        t_{m,i+1} - t_{m,i}\text{, for }n = m, \\
        \min\left[ t_{n,j} - t_{m,i} \right]_{t_{n,j} > t_{m,i}}\text{, for }n \neq m,
        \end{array}
    \right.
\end{align*}
This set is noted $\mathbf{D}_{m,i} = \left\{\Delta t_{m,i}^{n,-}, \Delta t_{m,i}^{n,+} \: \forall m \in \{1,\ldots,M\}\right\}$ and contains $2M$ different values, as illustrated in \autoref{fig:newmethod}. For ``border'' events (i.e., first and last events of any time window), the events right before/after the search time window are used as fill values (or the related value may be set to $\infty$, which would have no negative impact on the final sensitivity).

From the discussion above, if data is well described by a Poisson background with rate $r_{\rm bkg}$, it is expected that $\mathbf{D}_{m, i}$ values follow an exponential distribution with $\lambda = 1/r_{\rm bkg}$. This check may be formalized by computing, for each event $(m, i)$,
\begin{equation}
    \textrm{TS}_{m,i} = - \sum_{\Delta t \in \mathbf{D}_{m,i}} \log\left[ 1-\exp(-r_{\rm bkg} \Delta t) \right].
\end{equation}
The test statistic $\textrm{TS}$ follows an Erlang distribution ($\textrm{Erlang}(x; k) = x^{k-1} e^{-x} / (k-1)!$) with $k=2M$. A right tail might appear for background events that would accidentally fall close to where signal clusters are located, as illustrated in the right panel of \autoref{fig:newmethod}. This method is called \peanuts in the following, for Poisson-expectation Anomaly for Neutrino Transient Source Search.

As opposed to the MLE and PCA approaches, one value per event is computed instead of one per time window. It is then necessary not only to define a threshold $\textrm{TS}_{\rm thr}$ to select interesting events, but also to choose the minimum number of events $N_{\min}$ passing this threshold in a given time window to characterize an interesting excess. The $3\sigma$ threshold $\textrm{TS}_{\rm thr}$ can be found using the Erlang inverse survival function. The cut $N_{\min}$ is then useful to reduce the impact of the right tail mentioned in the above paragraph; $N_{\min} = 2$ is sufficient to remove all such background events.

\section{Results}
\label{sec:results}

To compare the three methods presented in \autoref{sec:method}, a toy scenario is considered :
\begin{itemize}
    \itemsep-2pt
    \item Background rate is fixed to \qty{20}{\milli\hertz}, representative of the expected background in IceCube GeV neutrino sample~\cite{IceCube:2021jwt} taken as example.
    \item The injected signal in a given time window is characterized by a strength $n_S$ that is the Poisson mean expected value. A Gaussian temporal profile, with mean $T_0$ and sigma $\sigma_T$, is considered. The typical values used for $T_0$ and $\sigma_T$ are \qty{0}{\second} and $0.5/2/\qty{5}{\second}$, respectively.
    \item A total of $M=200$ time windows is considered, and a signal is injected in $20\%$ of them.
\end{itemize}

\begin{figure}[hbtp]
    \centering
    \includegraphics[width=\linewidth]{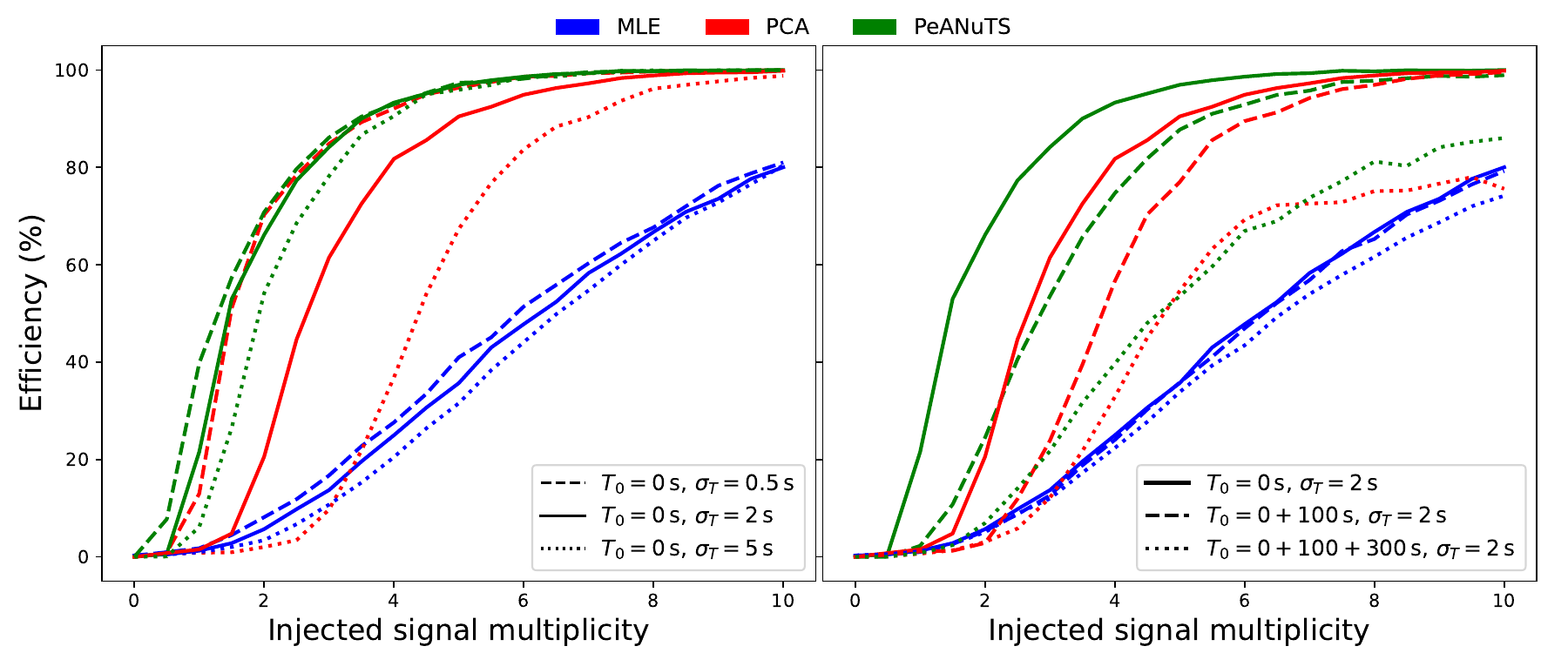}
    \caption{Comparison of the signal selection efficiencies for the three methods (MLE in blue, PCA in red, \peanuts in green) varying $n_S$ from $0$ to $10$, for the toy scenario with $T_0=\qty{0}{\second}$ (left) and for the two additional scenarios involving multiple signals (right). In the left panel, the line styles correspond to different values for $\sigma_T$. In the right panel, they are used to distinguish the different sets of $T_0$ values.}
    \label{fig:efficiencies}
\end{figure}

The comparison is then performed by simulating 10000 times such a scenario, referred to as pseudo-experiments. The efficiency of a given method is then evaluated as the fraction of time windows over all pseudo-experiments that are above the $3\sigma$ threshold as defined in \autoref{sec:method} for each method. For the new TS approach, the value of $m_{\min}$ is set to $2$. The left panel of \autoref{fig:efficiencies} shows the results as a function of the injected signal strength, for different signal widths. 

As expected, the MLE method is relatively robust with respect to the value of $\sigma_T$, as there is a corresponding fit parameter. However, as this approach only considers time windows independently, it fails to achieve high efficiencies, especially for a low number of events, this is also shown by the presence of a long tail on the signal distribution in \autoref{fig:stdmethod}. The PCA method is performing very well for short signals, where it is easier to identify a principal component within the dataset $\mathcal{D}$. For $\sigma_T < \qty{1}{\second}$, the latter may simply be the number of events in the bin corresponding to the signal window. For longer signals, the performance worsens significantly. Then, the \peanuts approach proposed here generally outperforms both MLE and PCA. Though it also has the same preference towards small signals, it is still efficient for larger $\sigma_T$. Compared to the two other methods, the signal strength threshold (corresponding to $>50\%$ selection efficiency) is lowered by factor $1.2$ to $3$ which, in turn, would correspond to a similar improvement in the flux constraints.

Two additional scenarios are then considered to cover cases where several timings are involved:
\begin{enumerate}[(a)]
    \item 20\% of the time windows have $T_0 = \qty{0}{\second}$, and another 20\% have $T_0 = \qty{100}{\second}$.
    \item 20\% of the time windows have $T_0 = \qty{0}{\second}$, 20\% have $T_0 = \qty{100}{\second}$, and 20\% have $T_0 = \qty{300}{\second}$.
\end{enumerate}
In both cases, the rest of the parameters do not change. For the PCA approach, as multiple clusters are expected to form for these new scenarios, the two first components are used (instead of only the first one) and the threshold is based on the bi-dimensional z-score of a point with respect to the background expected distribution (obtained from $\mathcal{D}_{\rm bkg}$). The related efficiencies are shown in the right panel of \autoref{fig:efficiencies}. The MLE results are relatively insensitive to these different scenarios as $T_0$ corresponds to a free parameter in the fit. On the contrary, the performance of the two other methods degrades significantly, as the signal is less precisely defined and therefore more difficult to identify. Nevertheless, the \peanuts method is still favored with respect to MLE and PCA.

\section{Summary and outlooks}
\label{sec:outlooks}

Two common techniques using a Maximum Likelihood Estimator or Principal Component Analysis are compared with a new approach called \peanuts in the context of a search for neutrino emission from subpopulations of transient sources. The latter achieves higher efficiencies as it strongly benefits from the potential correlations of signal timings in several time windows.

The new method may be used for searches in conservative time windows meant to catch any precursor, prompt, or slightly delayed signal. Notably, it may strongly enhance the discovery potential for neutrino samples for which the background rate is not low enough to allow claiming a discovery as soon as one or very few events are observed in this window. For instance, it is relevant for GeV neutrinos at large Cherenkov telescopes, as such samples are dominated by optical noises and by irreducible atmospheric neutrinos in this energy range. Moreover, the low number of Cherenkov photons emitted by GeV neutrinos does not allow for any precise direction reconstruction with current methods, limiting the possibility to benefit from using neutrino-source spatial correlations. This is then suited for the IceCube GeV neutrino sample~\cite{IceCube:2021jwt} or for the under-development similar sample in KM3NeT~\cite{Mauro:2023icrc}.

This method may be applied for binary mergers as detected with gravitational waves or for gamma-ray bursts seen throughout the electromagnetic spectrum. For both these categories, large catalogs (e.g.,~\cite{LIGOScientific:2021djp,grbweb}, respectively) are already available and may therefore be employed. For these objects, the existence and timing of the neutrino emission are currently unknown as it has never been observed, making it an interesting exploratory field.

As the \peanuts test statistic is quite general and only relies on arrival times, it could also be applied well beyond the scope presented here. Furthermore, adjustments may be possible to cover for potential non-Poissonian effects in the data or to include additional inputs in the analysis.

\subsection*{Acknowledgments}
M.L. is a Postdoctoral Researcher of the Fonds de la Recherche Scientifique - FNRS.

\bibliographystyle{JHEP}
\bibliography{references}

\end{document}